\documentclass[12pt]{article}

\usepackage{epsfig}

\newcommand{\be}{\begin{equation}}
\newcommand{\ee}{\end{equation}}
\newcommand{\bea}{\begin{eqnarray}}
\newcommand{\eea}{\end{eqnarray}}

\newcommand{\Dd}{{\mathcal D}}
\newcommand{\Pp}{{\mathcal P}}
\newcommand{\Bb}{{\mathcal B}}

\newcommand{\Qq}{{\mathcal Q}}
\newcommand{\pd}{\partial}

\newcommand{\ssim}{\mathop{\sim}\limits}

\begin{document}
\title{
\begin{flushright}
{\small SMI-15-99 }
\end{flushright}
\vspace{1cm}Two-Loop Diagrams in Noncommutative $\varphi^4_4$
theory}
\author{
I. Ya. Aref'eva${}^{\S}$, D.M. Belov${}^{\dag}$ and A.S.
Koshelev${}^{\dag}$ \\
\\${}^{\S}$ {\it Steklov Mathematical Institute,}\\ {\it Gubkin st.8,
Moscow, Russia, 117966}\\ arefeva@genesis.mi.ras.ru\\
\\${}^{\dag}$
{\it Physical Department, Moscow State University, }\\ {\it
Moscow, Russia, 119899} \\ belov@orc.ru, kas@depni.npi.msu.su}

\date {~}
\maketitle
\begin{abstract}
Explicit two-loop calculations in  noncommutative $\varphi^4_4$
theory are presented. It is shown that the model is two-loop
renormalizable.
\end {abstract}
%%%%%%%%%%%%%%%%%%%%%%%%%%%%%%%%%%%%%%%%%%%%%%%%%%%%%%%%%%%%%%%%%%%%%%%%%%%

\section{Introduction}

Noncommutative geometry\cite{book} and quantum groups \cite{Tax}
are of relevance to quantization of space-time (see, for example,
\cite{Mad,WZ,Luk} and references therein). Another possibility of
an application of   noncommutative geometry deals with  fields
that take values on q-deformed "manifolds", in particular on
quantum planes \cite{WZ} or on quantum groups \cite{AVqggf}. These
two approaches  are closely related. For example, gauge theory on
noncommutative torus is equivalent to noncommutative gauge theory
on the commutative torus. One of motivations to consider
noncommutative field  theories  is a hope that it would be
possible to cure quantum field theory
divergences\cite{AVqp,Mad,Kempf}.

The renovation of the interest in noncommutative field theories
has been stimulated by the paper of Connes, Douglas and Schwarz
\cite{CDS}, where it was shown that supersymmetric gauge theory on
noncommutative torus is naturally appeared in compactification of
Matrix theory \cite{BFSS} (see  \cite{NCYM} for further
developments). The appearance of noncommutative geometry in string
theory with a nonzero B-field \cite{CDS,Bfield,SW} and explicit
construction of a change of variables that shows an equivalence
between ordinary gauge fields \cite{SW} and noncommutative gauge
fields  raises  a question about selfconsistency of noncommutative
Yang-Mills (NCYM) theory. NCYM theory is related with deformation
quantization, see  \cite{AV}. In such type of models ultraviolet
divergences are still present \cite{filk}. Moreover,
renormalizability is  not evident for noncommutative field
theories. Nonrenormalizability of NCYM theory would mean
inconsistency of a string theory in the B-field background at
least at the zero-mode level. Therefore, it is crucial from string
point of view to clarify this question. Explicit calculations
performed at one-loop level show renormalizability in this
approximation \cite{ren}. The next orders have not  been
calculated yet. About a general discussion of renormalizability
see \cite{Ch}.

The goal of this paper is  to show  renormalizability of the
noncommutative scalar theory  in two-loop approximation. We will
construct explicitly one and two-loop counterterms
 and show that renormalized
1PI functions are well-defined  for non-exceptional momenta
(compare with IR finiteness of renormalized 1PI functions in
massless theories). The similar calculations for NCYM are in
progress. Note that there are more similarity between calculations
of counterterms in NCYM theory and those  in charged scalar field
theory \cite{AV}.

The paper is organized as follows. In Sect.2 we formulate the
model and explain why we will work in the language of single-line
graphs and symmetric vertices. In Sect.3 we present the one loop
calculations and in Sect.4 we schematically present two-loop
calculations. Details of these calculations are collected in
Appendix. In the last Section we  note a problem with IR
divergences for higher loops renormalized 1PI functions (we have
this problem in spite of we deal with a massive theory).

\section{The Model}
We consider the theory with the action
\be
\label{theory} S=S_0+S_{int}=\int d^dx\,[
\frac{1}{2}(\pd_{\mu}\varphi)^2 + \frac{1}{2}m^2\varphi^2 + \frac
g{4!}(\varphi \star\varphi \star\varphi \star\varphi)(x)],
\ee
where $\star$ is a Moyal product  $(f\star
g)(x)=e^{i\xi\theta^{\mu\nu}\pd_{\mu}\otimes\pd_{\nu}}f(x)\otimes
g(x)$, $\xi$ is a deformation parameter, $\theta^{\mu\nu}$ is a
non-degenerate skew-symmetric real constant matrix, $\theta^2=-1$,
$d$ is even. $p_1\wedge p_2\equiv\xi p_1\theta p_2$. Let us
rewrite the interaction in the Fourier components,
\be
\label{int}
S_{int}=\frac{g}{4!(2\pi)^d}\int dp_1dp_2dp_3dp_4\, e^{-i p_1\wedge
 p_2-i p_3\wedge p_4}
\varphi(p_1)\varphi(p_2)\varphi(p_3)\varphi(p_4)\delta(p_1+p_2+p_3+p_4).
\ee
We see the following distinguish features of the deformed theory
as compared with standard local $\varphi^4_d$ model:
\begin{itemize}
\item
There are non-local phase factors in the vertex.
\item
These factors provide regularization for some loop integrals but
not for all \cite{ren,AV,SB}.
\item
To have renormalizability  the sum of divergences in each order of
perturbation theory must have a phase factor already present in
the action.
\end{itemize}

To single out phase factors it is convenient  to use the 't Hooft
double-line graphs and a notion of planar graphs. For planar
graphs the  phase factors do not affect the Feynman integrations
at all \cite{filk}. In particular the planar graphs have exactly
the same divergences as in the commutative theory\cite{SB}. There
are no superficial divergences in nonplanar graphs since they are
regulated by the phase factors \cite{ren,AV,SB,Ch}. Moreover,
oscillating phases regulate also divergent subgraphs, unless they
are not divergent planar subgraphs.

\begin{figure}[h]
\begin{center}
\epsfig{file=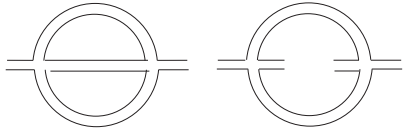,
   width=150pt,
  angle=0,
 }
\caption{Planar graph with a nonplanar subgraph}
\label{F1}
\end{center}
\end{figure}

So, at first sight  it seems that the proof of renormalizability
is rather trivial. One has divergences only in planar graphs, so
one can use the fact that planar theory is renormalizable if its
scalar counterpart does \cite{Hooft}. Or in other words, one can
take the planar approximation of ordinary theory, find divergences
within this approximation and write  counterterms in
noncommutative theory as  divergent parts of planar graphs
multiplied  on the phase factors. However these arguments work
only for superficial divergences. The situation is more subtle for
divergent subgraphs. The reason is that a planar graph can contain
nonplanar subgraphs (see  simple example on Fig.\ref{F1}) and
these divergences should be also removed. Therefore,
renormalizability of the theory (\ref{theory}) is not obvious.

\begin{figure}[h]
\begin{center}
\epsfig{file=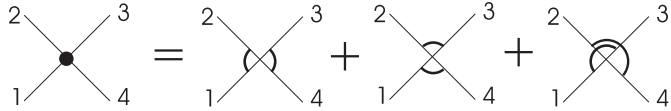,
   width=250pt,
  angle=0,
 }
\caption{The symmetric vertex}
\label{F2}
\end{center}
\end{figure}

In explicit calculations we will use single-line graphs and
symmetric vertices. After symmetrization of (\ref{int}) we get
\begin{eqnarray}
S_{int}=\frac{g}{3\cdot 4!}\frac{1}{(2\pi)^d}\int
dp_1dp_2dp_3dp_4\,
\varphi(p_1)\varphi(p_2)\varphi(p_3)\varphi(p_4)\delta(p_1+p_2+p_3+p_4)
\nonumber\\ \times\left[\cos(p_1\wedge p_2)\cos(p_3\wedge
p_4)+\cos(p_1\wedge p_3) \cos(p_2\wedge p_4) +\cos(p_1\wedge
p_4)\cos(p_2\wedge p_3)\right]
\end{eqnarray}
and the vertex is a sum of three terms (see Fig.\ref{F2}).

All further examination is confined to the case $d=4$.

\section{One Loop Counterterms}
In this section we will compute explicitly  one-loop counterterms
using dimensional regularization $d=4-2\epsilon$. We will also
present the explicit form of finite part for two point and four
point 1PI functions, $\Gamma ^{(2)}$ and $\Gamma ^{(4)}$, in the
one loop approximation. We use the standard notations for
perturbation expansion of 1PI-functions, $\Gamma ^{(i)}=\sum _n
g^n\Gamma _{n}^{(i)}$, $\Gamma ^{(2)}=\Gamma
_{f.p.}^{(2)}+\Delta\Gamma ^{(2)} $ and $\Gamma ^{(4)}=\Gamma
_{f.p.}^{(4)}+\Delta \Gamma ^{(4)}$
\begin{figure}[h]
\begin{center}
\epsfig{file=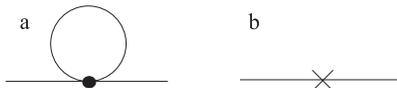,
   width=150pt,
  angle=0,
 }
\caption{$\Gamma _{1}^{(2)}-\Delta\Gamma _{1}^{(2)}$} \label{F3}
\end{center}
\end{figure}

The only graph \ref{F3}a contributes to $\Gamma _{1}^{(2)}$ and $$
\Gamma _{1}^{(2)}=-\frac{g}{6(2\pi)^d} \int dk\frac{2+\cos
2p\wedge k}{k^2+m^2}= $$
\be
\frac{g}{32\pi^2}\frac23
\left[m^2\left(\frac{1}{\epsilon}+\psi(2)-\ln\frac{m^2}{4\pi\mu^2}\right)-
\sqrt{\frac{m^2}{\xi^2p^2}}K_1(2m\xi |p|)\right]. \label{gamma2}
\ee
Its divergent part is subtracted by the counterterm \ref{F3}b.
Note that the representation (\ref{gamma2}) takes place only for
$p\neq 0$.

\begin{figure}[h]
\begin{center}
\epsfig{file=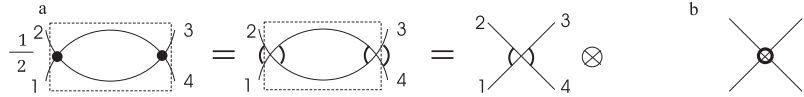,
   width=300pt,
  angle=0,
 }
\caption{a) $\Delta\Gamma _{2,s}^{(4)}$ $\quad$ b) Cross denotes
the 4-vertex counterterm} \label{F4}
\end{center}
\end{figure}

$\Gamma _{2}^{(4)}$ is a  sum of s-, t- and u-channel graphs,
$\Gamma _{2}^{(4)}=\Gamma _{2,s}^{(4)}+ \Gamma _{2,t}^{(4)}+\Gamma
_{2,u}^{(4)}$. The explicit form of $\Gamma _{2,s}^{(4)}$ is
\be
   \label{7a}
\Gamma_{2,s}^{(4)}=\frac{g^2}{18(2\pi)^{d}}\int
\frac{\Pp_{\ref{F4}a}(\{p\},k)}{(k^2+m^2)((P+k)^2+m^2)} dk
\ee
where $P=p_1+p_2$, $\Pp_{\ref{F4}a}(\{p\},k)=[2\cos(p_1\wedge
p_2)\cos(k\wedge P)+\cos(p_1\wedge p_2+(p_1-p_2)\wedge k)]\times$
$[2\cos(p_3\wedge p_4)\cos(k\wedge P)+\cos(p_3\wedge
p_4+(p_4-p_3)\wedge k)] $. The trigonometric polynomial
$\Pp_{\ref{F4}a}$  can be rewritten in the form
\be
           \label{tp7a}
\Pp_{\ref{F4}a}=2\cos(p_1\wedge
p_2)\cos(p_3\wedge p_4)+
\sum _{j} {}' c_j e^{i\Phi^j(p)+ib^j(p)\wedge k}
\ee
where the sum $\sum '$ goes over all $j$ for which linear
functions $b^j(p)$ are nonzero for almost all $\{p\}$. The only
first term in  (\ref{tp7a}) contributes to a pole part (see Fig.
\ref{F4}a) and we have
$$
\Gamma _{2,s}^{(4)}=\frac{g^2}{32\pi^2}\frac29
\left[\cos(p_1\wedge p_2)\cos(p_3\wedge p_4)\left(\frac
1{\epsilon}+\psi(1)-\int_0^1dx\ln\frac{m^2+x(1-x)P^2}{4\pi\mu^2}
\right) +\right.$$
\be
\label{rep4} +\left.\sum_j {}'c_j e^{i\Phi^j(p)}
\int_0^1dxK_0(\sqrt{\xi^2{b^{j}}^2(m^2+P^2x(1-x))}) e^{ixb^j\wedge
P}\right].
\ee
The representation (\ref{rep4}) is well defined only if all
$b^{j}$ are nonzero, i.e. for non exceptional momenta. It is the matter
of simple algebra to sum up the divergent parts of $\Gamma
_{2,s}^{(4)}$,  $\Gamma _{2,t}^{(4)}$ and $\Gamma _{2,u}^{(4)}$ to
obtain graph with symmetric vertex.

Therefore, in the one-loop we have
\be
\Delta \Gamma_{1l}^{(2)}= \frac{g}{48\pi^2}\frac{m^2}{\epsilon}
\label{21l}
\ee
\be
\Delta \Gamma_{1l}^{(4)} =\frac{g^2}{16\pi^2}
\frac{1}{9\epsilon}[\cos(p_1\wedge p_2)\cos(p_3\wedge p_4)+ \cos(
p_1\wedge p_3)\cos(p_2\wedge p_4) +\cos(p_1\wedge p_4)\cos(
p_2\wedge p_3)] \label{41l}\ee

\section{Two Loop Counterterms}
Let us consider $\Gamma _{2}^{(2)}$. The corresponding graphs are
shown in Fig. \ref{F5}. Graph \ref{F5}a is compensated by graph
\ref{F5}c (see Appendix B).

\begin{figure}[h]
\begin{center}
\epsfig{ file=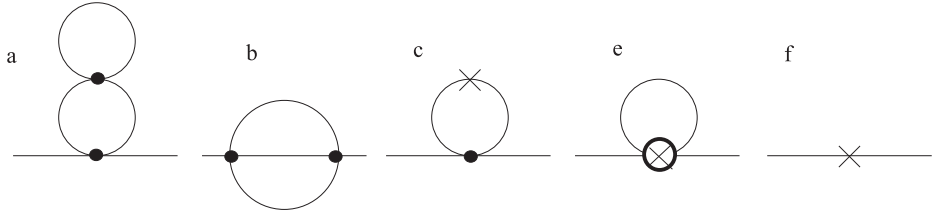,
   width=350pt,
  angle=0,
 }
\caption{Graphs contributing  to  $\Gamma _{2}^{(2)}$}
\label{F5}
\end{center}
\end{figure}

Let us consider graph \ref{F5}b (an explicit expression for this
graph is presented in Appendix C, eq.(\ref{g5b})). This graph has
superficial divergence which is removed by a local counterterm.
Also it has divergent subgraphs. Considering one of them (Fig
\ref{F6}a) ($b_{12}=0, b_1=0, b_2\neq 0$, i.e. case i) in Appendix
C) we see that a contribution of this divergent subgraph can be
represented as \ref{F6}d which is nothing but a non-local tadpole
with a one-loop renormalized coupling constant. This shows the
two-loop  renormalizability of $\Gamma^{(2)}$.

\begin{figure}[h]
\begin{center}
\epsfig{ file=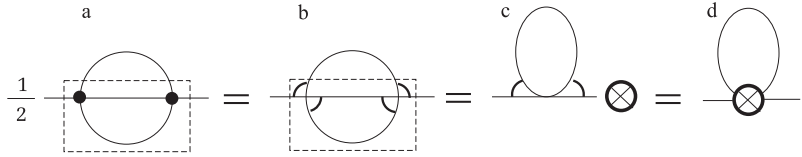,
   width=300pt,
  angle=0,
 }
\caption{Divergent subgraph of  \ref{F5}b}
 \label{F6}
\end{center}
\end{figure}

\begin{figure}[h]
\begin{center}
\epsfig{ file=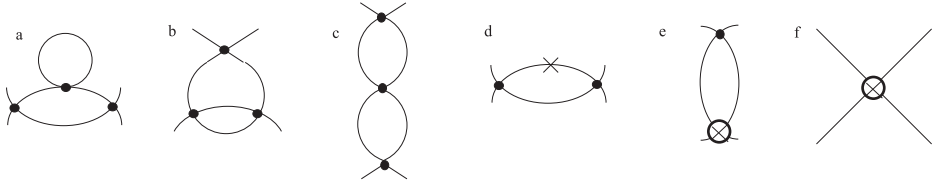,
   width=350pt,
  angle=0,
 }
\caption{a), b) and c) show graphs contributing  to  $\Gamma
_{3}^{(4)}$; d), e) and f) show graphs with counterterms (cross
denotes counterterm (\ref{21l}))} \label{F7}
\end{center}
\end{figure}

Let us now prove two-loop renormalizability of $\Gamma ^{(4)}$. We
have three two-loop graphs (\ref{F7}a,b,c) and corresponding
crossing graphs. In the same order of $g$ there are also graphs
with counterterms, (\ref{F7}d,e,f).

\begin{figure}[h]
\begin{center}
\epsfig{ file=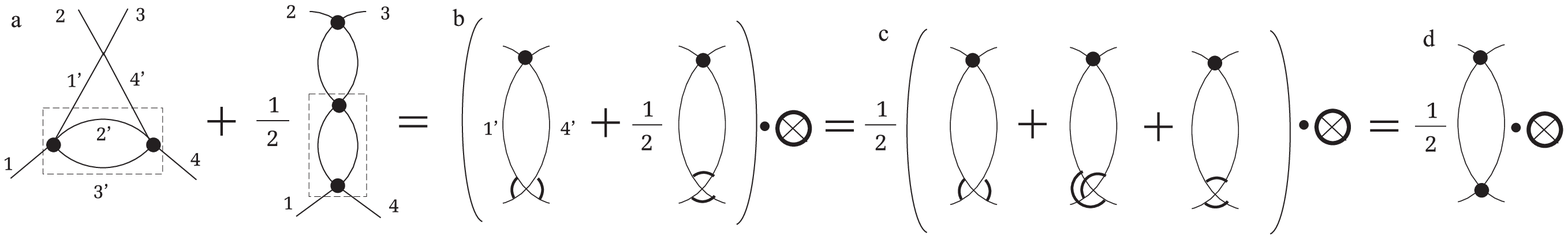,
   width=400pt,
  angle=0,
 }
\caption{Divergent parts of graphs \ref{F7}b and \ref{F7}c}
\label{F8}
\end{center}
\end{figure}

Graph \ref{F7}a has only one divergent subgraph and this
divergence is compensated by graph \ref{F7}d.  More interesting is
a compensation of divergences of graphs \ref{F7}b and \ref{F7}c
caused by one-loop divergent subgraphs.  As we will see these
divergences are compensated by \ref{F7}e only if we take the sum
of graphs \ref{F7}b and \ref{F7}c. Indeed, we have equalities
presented in fig.\ref{F8}. The second equality in fig.\ref{F8} is
due to symmetry under exchange $1' \leftrightarrow 4'$. So, we
have a compensation of the terms presented in fig.\ref{F7}e  and
\ref{F8}d. This compensation is checked explicitly in Appendix.

Therefore, in two loops we have
\be
\Delta \Gamma_{2l}^{(2)} = \pi^4 \frac{g^2}{3^2(2\pi)^8}
\left[\frac {5m^2}{8\epsilon^2}-\frac
{3m^2}{4\epsilon}(\frac32+\Psi(1)-\ln\frac{m^2}{4\pi\mu^2})-\frac{p^2}{16\epsilon}\right]
\ee
$$ \Delta \Gamma_{2l}^{(4)} = \frac{g^3}{3^3(2\pi)^8} \pi^4
\left[\frac 1{\epsilon^2}-\frac 1{2\epsilon}\right]
\left[\cos(p_1\wedge p_2)\cos(p_3\wedge p_4)+\right. $$
\be
+\left.\cos(p_1\wedge p_3)\cos(p_2\wedge p_4)+ \cos(p_1\wedge
p_4)\cos(p_2\wedge p_3)\right]
\ee

\section{Discussion and Conclusion}

Let us compare our counterterms with corresponding counterterms in
the ordinary theory. In one loop for $\Delta \Gamma_{1l}^{(4)}$ we
have extra factor $2/9$ as compared with the local $\varphi ^4_4$
theory. "2" comes from free terms of the trigonometric polynomial
(\ref{tp7a}) and 1/9 is from the vertex. We have the factor $2/9$
also for the one-loop $4$-point planar graph. Relative factors of
divergent parts of two-loop graphs are collected in the Table.

{
\renewcommand{\arraystretch}{2.5}
$$
\begin{tabular}{|c|c|c|c|}
\hline
Graph  & Noncommutative & Commutative & Planar \\
\hline
\epsfig{file=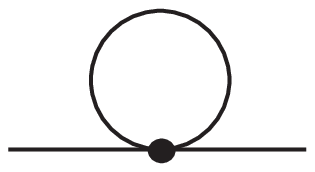,
   width=30pt,
  angle=0,
 }
   & 1/3 & $1/2$  & $1/3$ \\
\hline
\epsfig{file=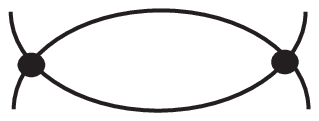,
   width=30pt,
  angle=0,
 }
   & 1/9 & $1/2$ & $1/9$ \\
\hline
\epsfig{file=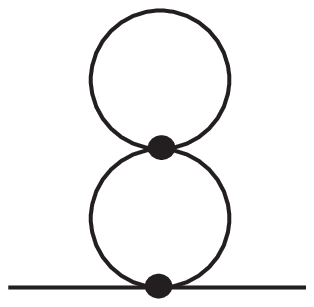,
   width=30pt,
  angle=0,
 }
   & 1/9 &$1/4$  & $1/9$ \\
\hline
\epsfig{file=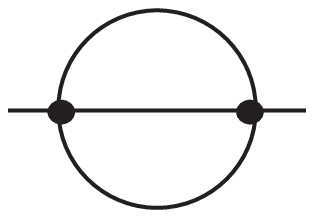,
   width=30pt,
  angle=0,
 }
   & 1/36 & $1/6$ & $1/36$ \\
\hline
\epsfig{file=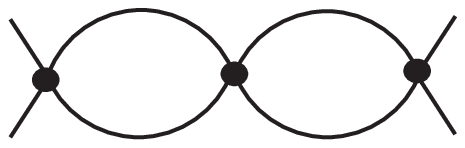,
   width=30pt,
  angle=0,
 }
   & $1/(27\cdot 2)$ & $1/4$ & $1/(27\cdot 2)$ \\
\hline
\epsfig{file=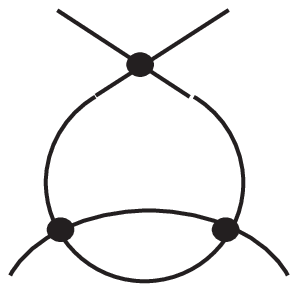,
   width=30pt,
  angle=0,
 }
   & $1/(27\cdot 2)$ & $1/2$ & $1/(27\cdot 2)$ \\ \hline
\end{tabular}
$$
}

It is obvious that our renormalized 1PI functions do not admit the
limit $\xi \to 0$. A  behaviour of $\Gamma ^{(2)}_1$ for $p^2 \to
0$ is the same as its  behaviour for $\xi \to 0$. We have
\be
\label{IR} \Gamma ^{(2)}_{1,f.p.} \ssim_{p^2 \to 0} \frac{c}{(\xi |p|)^2}
\ee
Caused by this asymptotic there are problems with a IR behaviour
of graphs with tadpoles. They produce divergence in the IR region
if the number of insertions is more then $n>2$ (compare with an
example of ref.[21] in \cite{Ch}).

Perhaps, it would be also interesting to develop the theory of
noncommutative quantum (gauge) fields not only on the torus but
also on  more general manifolds, in particular on  Poisson
manifolds.  Perhaps the question of renormalizability of
noncommutative quantum theory on  certain Poisson manifolds will
get a more favourable resolution \cite{AV}.

\section*{Note added}
After the finishing this paper we became aware
of the work by  Shiraz Minwalla, Mark Van Raamsdonk and  Nathan
Seiberg, hep-th/9912072, where  mixing of the UV and the IR is
discussed.

%%%%%
\section*{Acknowledgments}

We would like to thank P. B. Medvedev, O.A. Rytchkov  and I.V.
Volovich for useful discussions. This work was supported in part
by RFFI grant 99-01-00166  and by grant for the leading scientific
schools 96-15-96208. I.A. is also supported by INTAS grant
96-0698.
%\newpage

{\small
\section*{Appendix}
\renewcommand {\theequation}{\thesection.\arabic{equation}}

\appendix
\setcounter{equation}{0}
\section{Notations}
We  denote external momenta by $p_i$ and loop momenta by $k_1$ and
$k_2$. Let $\Pp_{\Gamma}(\{p\},\{k\})$ be a trigonometric
polynomial corresponding to a two-loop graph $\Gamma$. It can be
represented in a form
\be
\label{pol}
\Pp_{\Gamma}=\sum_{j}c_j\exp[i\Phi^j(p)+ i {b^j_1}
\wedge k_1+ib^j_2 \wedge k_2+2i b^j_{12} k_1\wedge k_2],
\ee
where $\Phi^j(p)$ are functions of external momenta, $b^j_{i}$,
$i=1,2$ are linear functions of $p_i$, $b^j_{12}$ and $c_j$ are
real numbers. We denote $\Bb_j(p_i)=c_j\exp[i\Phi^j(p)]$. Each
term of (\ref{pol}) gives a contribution to
  $\alpha$-representation of  a two-loop diagram as
$$ \int
e^{[-(a_1k_1^2+a_2k_2^2+2a_{12}k_1k_2+l_1'k_1+l_2k_2+M^2)+
   i b_1 \wedge k_1+i b_2\wedge k_2+2i b_{12}k_1\wedge  k_2]}
   d^dk_1d^dk_2\prod _i d\alpha _i=
$$
$$\int  \frac{\pi^d}{\Dd^{d/2}}
 \exp[-M^2+\frac{a_1l_2^2+a_2l_1^2-2a_{12}l_1l_2-
\xi^2(a_1b_2^2+a_2b_1^2-2a_{12}b_1b_2)}
{4\Dd}+ $$
\be
i\xi\frac{a_1l_2\theta b_2+a_2l_1\theta b_1-a_{12}(l_1\theta b_2+
l_2\theta b_1)}{2\Dd}+ i\xi b_{12}\frac{(l_1-i\xi b_1\theta)
\theta(l_2+ i\xi\theta b_2)}{2\Dd}] \prod _i d\alpha _i
\label{2loop}
\ee
where $\Dd=a_1a_2-a_{12}^2+\xi^2b_{12}^2$,
$~a_1, a_2, a_{12}$ depend on $\alpha_i$, $l_1, l_2$ and $M$
depend on $\alpha_i$ and $p_j$, $b_1, b_2$ depend on $p_j$,
$b_{12}$ is a constant.

To single out divergences we will separate the sums over $j$ on
several groups. We say that

\begin{itemize}
  \item $j\in J(b_{12})$ if  $b^j_{12}=0$
 and denote
$\sum _{j\in J(b_{12})}=\sum _{b_1,b_2}$
  \item $j\in J(b_{12},b_{1})$ if  $b^j_{12}=0$ and $b_1=0$ and $b_2\neq
0$ (for almost all $p$) and denote $\sum _{j\in
J(b_{12},b_{1}))}=\sum _{b_2}$ (similarly for $1\leftrightarrow
2$.)
  \item $j\in J(b_{12},b_{1}-b_2)$ if  $b^j_{12}=0$ and  $b_1=b_2\neq 0$
(for almost all $p$) and denote $\sum _{j\in
J(b_{12},b_{1}-b_2))}=\sum _{b_1=b_2}$.
\end{itemize}

\section{Compensation of graphs \ref{F5}a and \ref{F5}c}

The graph \ref{F5}a has the following analytic expression $$
\Gamma_{\ref{F5}a}(p)=\frac14\left[\frac{g}{3(2\pi)^d}\right]^2\int
d^dk_1d^dk_2\frac{\Pp_{\ref{F5}a}(p,k_1,k_2)}
{(k_1^2+m^2)^2(k_2^2+m^2)}, $$
\be
                          \label{gr5a}
\Pp_{\ref{F5}a}(p,k_1,k_2)=[2+\cos(2k_1\wedge
k_2)][2+\cos(2k_2\wedge p)].
\ee
Considering terms with $b_{12}=0$ in (\ref{gr5a}) we obtain $$
\Delta\Gamma_{\ref{F5}a}(p)=-\left[\frac{g}{3
(2\pi)^4}\right]^2\pi^4\left(\frac{m^2}{\epsilon^2}+\frac{\Psi(2)+\Psi(1)}{\epsilon}+
\frac{m^2}{2\epsilon}
K_0(2\xi
m|p|)\right)\left[\frac{4\pi\mu^2}{m^2}\right]^{2\epsilon}. $$ For
graph \ref{F5}c we have $$
\Gamma_{\ref{F5}c}(p)=\left[\frac{g}{3(2\pi)^d}\right]^2
\frac{\pi^2m^2}{2\epsilon}\int d^dk\frac{2+\cos(2p\wedge k)}
{(k^2+m^2)^2}= \left[\frac{g}{3 (2\pi)^4}\right]^2\pi^4
\left(\frac {m^2}{\epsilon^2}+\frac{\Psi(1)}{\epsilon}+\frac {m^2}{\epsilon}K_0(2\xi
m|p|)\right) \left[\frac{4\pi\mu^2}{m^2}\right]^{\epsilon}.
$$ We
see that  $\Gamma_{\ref{F5}a}(p)+\Gamma_{\ref{F5}c}(p)$ is finite.

\section{Two-loop corrections to $\Gamma^{(2)}(p^2)$}
For Feynman graph \ref{F5}b we have $$
\Gamma_{\ref{F5}b}(p)=\frac16\left[\frac{g}{3
(2\pi)^d}\right]^2\int
\frac{d^dk_1d^dk_2\,\Pp_{\ref{F5}b}(p,k_1,k_2)}
{(k_1^2+m^2)(k_2^2+m^2)((k_1+k_2-p)^2+m^2)}, $$ where the
trigonometric polynomial $\Pp_{\ref{F5}b}(p,k_1,k_2)$ has a form
$$ \Pp_{\ref{F5}b}(p,k_1,k_2)=[1+\cos(2k_1\wedge
k_2)][1+\cos(2p\wedge(k_1+k_2))]+\frac{1}{2}[1+ \cos(2k_1\wedge
k_2+2p\wedge(k_1-k_2))]+ $$ $$ +\cos(2p\wedge k_1)+\cos(2p\wedge
k_2)+\cos(2k_1\wedge k_2+2p\wedge k_1) +\cos(2k_1\wedge
k_2-2p\wedge k_2). $$ Using $\alpha$-representation we write
$\Gamma_{\ref{F5}b}(p)$ as
\be
\label{g5b} \Gamma_{\ref{F5}b}(p)=
\frac{\pi^d}{6}\left[\frac{g}{3(2\pi)^d}\right]^2\sum_j\int_0^\infty
\frac{d\alpha_1d\alpha_2d\alpha_3 \Bb^j(p)e^{\Qq
(\alpha,b^j)}}{\Dd^{d/2}}, \quad
\Dd=(\alpha_1\alpha_2+\alpha_2\alpha_3+\alpha_3\alpha_1+\xi^2
{b^j_{12}}^2),
\ee
$$\Qq (\alpha,b)=-m^2(\alpha_1+\alpha_2+\alpha_3)-
p_1^2\frac{\alpha_1\alpha_2\alpha_3+ \alpha_2\xi^2b_{12}^2}{\Dd}
-p_1^2\xi^2\frac{\alpha_1b_2^2+
\alpha_3b_1^2+\alpha_2(b_1-b_2)^2}{4\Dd} $$ Integral (\ref{g5b})
may have divergences if $b_{12}=0$ and simultaneously  one of the
following conditions is realized i) $b_1=0,b_2\neq0$; ii)
$b_1\neq0, b_2=0$; iii) $b_1=b_2\neq0$; iv) $b_1=b_2=0$. We get
\be
\Delta \Gamma_{\ref{F5}b}(p)=\frac{C_{\ref{F5}b}}{\epsilon^2}+
\frac{C'_{\ref{F5}b}}{\epsilon}+
\left[\frac{g}{3(2\pi)^d}\right]^2\frac{\pi^4}
{\epsilon}\sqrt{\frac{m^2}{\xi^2p^2}}K_1(2\,\xi
m|p|).
\ee
For graph \ref{F5}e (see also fig. \ref{F6}d). we get
\be
\Delta\Gamma_{\ref{F5}e}(p)=\frac{C_{\ref{F5}e}}{\epsilon^2}+
\frac{C'_{\ref{F5}d}}{\epsilon}
-\left[\frac{g}{3(2\pi)^d}\right]^2\frac{\pi^4}
{\epsilon}\sqrt{\frac{m^2}{\xi^2p^2}}K_1(2\,\xi m|p|).
\ee
Hence we see that nonlocal divergences appeared in the
noncommutative theory are compensated in the sum
$\Gamma_{\ref{F5}b}(p)+\Gamma_{\ref{F5}e}(p)$.

\section{Two-loop corrections to $\Gamma^{(4)}(p)$}
The divergent part of Feynman graph  \ref{F7}a is exactly
compensated by the graph  \ref{F7}d. A nontrivial compensation
takes place for graphs \ref{F7}b, \ref{F7}c and \ref{F7}e. For
graph \ref{F7}b we have $$
\Gamma_{\ref{F7}b}(p_i)=-\frac12\frac{g^3}{3^3(2\pi)^{2d}}\int
\frac{d^dk_1d^dk_2\,
\Pp_{\ref{F7}b}(p_i,k_1,k_2)}{(k_1^2+m^2)(k_2^2+m^2)
((k_2+P)^2+m^2)((k_1+k_2- p_4)^2 +m^2)}= $$
\be
\label{d7b}
-\frac{\pi^d}{2}\frac{g^3}{3^3(2\pi)^{2d}}\sum_{b}\int_0^\infty
\frac{d\alpha_1d\alpha_2d\alpha_3d\alpha_4\,\Bb ^j(p_i,k_1,k_2)
e^{\Qq^j(\alpha,b^j)}}{\Dd^{d/2}} +...
\ee
Here dots denote the terms with $b_{12}\neq 0$ that are not
relevant to   our consideration,
$$
\Qq^j(\alpha,b)=-m^2[\alpha_1+\alpha_2+\alpha_3+\alpha_4]-
\frac{1}{\Dd}[P^2\alpha_2\alpha_3(\alpha_1+\alpha_4)+p_4^2\alpha_1\alpha
_2\alpha_4+ p_3^2\alpha_1\alpha_3\alpha_4]$$ $$
-\frac{\xi^2}{4\Dd}[(\alpha_2+\alpha_4)b_1^2+\alpha_1b_2^2+\alpha_4(b_1-
b_2)^ 2]-\frac{i\xi}{\Dd}[(\alpha_1\alpha_4
p_4-\alpha_3(\alpha_1+\alpha_4)P)\theta b_2+
\alpha_4(\alpha_2p_4-\alpha_3p_3)\theta b_1], $$ and
$\Dd=(\alpha_1\alpha_2+\alpha_1\alpha_3+
\alpha_1\alpha_4+\alpha_2\alpha_4+\alpha_3\alpha_4)$.
 The  integral in RHS of (\ref{d7b})
diverges only in the cases i) $b_1=0, b_2\neq 0$ and ii) $b_1=b_2=0$.
In the first case one gets
\be
\Delta \Gamma ^{i)}_{\ref{F7}b}=
-\frac{g^3}{3^3(2\pi)^{2d}}\frac{\pi^4}{\epsilon}
\sum_{b_2}
\int_0^1dx\,\Bb_{b_2}(p_i)K_0\left(\sqrt{\xi^2b_2^2[m^2+P^2x(1-x)]}\right)
\exp(i\xi xb_2\,\theta P).
\ee
To determine $\Bb_{b_2}$ we  pick out from $\Pp(p_i,k_1,k_2)$ a
trigonometric polynomial that has $b_1=0$. This polynomial (after
performing symmetrization $p_3\leftrightarrow p_4$, compare with
fig. \ref{F8}a,b) has a form $$
\Pp_{\ref{F7}b}^{b_1=0}(p_i)=[2\cos(p_1\wedge p_2)\cos(k_2\wedge
P)+\cos(p_1\wedge p_2+(p_1-p_2)\wedge k_2)] $$ $$
[\,\cos(p_3\wedge p_4)\cos(k_2\wedge P)+\cos\,(p_3\wedge
p_4+(p_4-p_3) \wedge k_2)]. $$ $\Delta \Gamma ^{ii)}_{\ref{F7}b}$
has divergences as in the local theory,
\be
\Delta \Gamma ^{ii)}_{\ref{F7}b}=-\frac{\pi^4}{2}
\frac{g^3}{3^3(2\pi)^8} \cos(p_1\wedge p_2)\cos(p_3\wedge p_4)
\left[\frac{1}{2\epsilon^2}+\frac{1}{2\epsilon}+\frac{\Psi(1)}{\epsilon}-
\frac{1}{\epsilon}\int_0^1 dx\ln
\frac{m^2+P^2x(1-x)}{4\pi\mu^2}\right].
\ee

Turning to  graph  \ref{F7}c we have $$
\Gamma_{\ref{F7}c}(p)=-\frac14\frac{g^3}{3^3(2\pi)^{2d}}\int
\frac{d^dk_1d^dk_2\,
\Pp_{\ref{F7}c}(p_i,k_1,k_2)}{(k_1^2+m^2)((P+k_1)^2+m^2)
(k_2^2+m^2)((k_2-P)^2 +m^2)} $$ Using the same technique as in the
previous calculations we see that the graph has divergences in the
cases i) $b_1=0,b_2\neq 0$, ii) $b_1\neq0,b_2=0$ and iii)
$b_1=b_2=0$ simultaneously with $b_{12}=0$ (in all cases). For
example,
\be
\label{g7cii}
\Delta\Gamma^{ii)}_{\ref{F7}c}(p)=-\frac{g^3}{3^3(2\pi)^2}
\frac{\pi^4}{2\epsilon}
\sum_{ b_1}
\int_0^1dx\,\Bb_{b_1}(p_i)K_0\left(\sqrt{\xi^2b_1^2[m^2+P^2x(1-x)]}\right)
\exp(i\xi xb_1\,\theta P).
\ee
To determine $\Bb_{b_1}(p)$ we pick out from
$\Pp_{\ref{F7}c}(p_i,k_1,k_2)$
a trigonometric polynomial that has
$b_2=0$
$$ \Pp^{b_2=0}_{\ref{F7}c}=2\cos(p_3\wedge
p_4)\cos(k\wedge P)[2\cos(p_1\wedge p_2) \cos(k\wedge
P)+\cos(p_1\wedge p_2+(p_1-p_2)\wedge k)].
$$
$\Delta\Gamma^{ii)}_{\ref{F7}c}(p)$
is obtained from (\ref{g7cii})
by $(1,2) \to (3,4)$. $\Delta \Gamma ^{iii)}_{\ref{F7}c}$ has
divergences as in the local theory,
\be
\Delta \Gamma ^{iii)}_{\ref{F7}c}=
-\frac{\pi^4}{2}\frac{g^3}{3^3(2\pi)^8}
\cos(p_1\wedge p_2)\cos(p_3\wedge p_4)
\left[\frac{1}{\epsilon^2}+\frac{2\Psi(1)}{\epsilon}-
\frac{2}{\epsilon}\int_0^1 dx\ln \frac{m^2+P^2x(1-x)}{4\pi\mu^2}\right].
\ee

Finally,  graph \ref{F7}e yields
 $$
\Gamma_{\ref{F7}e}(p_i)=
\frac{g^3}{3^3(2\pi)^{2d}}\frac{\pi^2}{2\epsilon} \int
\frac{d^dk\,\Pp_{\ref{F7}e}(p_i,k)}{(k^2+m^2)((k+P)^2+m^2)}=
$$
$$
\pi^4\frac{g^3}{3^3(2\pi)^8} \cos(p_1\wedge
p_2)\cos(p_3\wedge p_4)
\left[\frac{1}{\epsilon^2}+\frac{\Psi(1)}{\epsilon}-
\frac{1}{\epsilon}\int_0^1 dx\ln
\frac{m^2+P^2x(1-x)}{4\pi\mu^2}\right]
 $$
 $$ \frac{\pi^4}{\epsilon}\frac{g^3}{3^3(2\pi)^{2d}}
\sum_{b}\int_0^1dx\,\Bb_b(p_i)K_0
\left(\sqrt{\xi^2b^2[m^2+P^2x(1-x)]}\right)
\exp(i\xi xb\,\theta P). $$

$\sum_{b}$ means that the sum is over $j$
 with $b^j \neq 0$ (see (\ref{tp7a})). The $\Bb_b(p_i)$ are coming from
$$ \Pp_{\ref{F7}e}(p_i,k)=[2\cos\,p_1\wedge p_2\cos k\wedge
P+\cos(p_1\wedge p_2+(p_1-p_2)\wedge k)] $$ $$
\times[2\cos\,p_3\wedge p_4\cos k\wedge P+\cos(p_3\wedge
p_4+(p_4-p_3)\wedge k)]+ k)]. $$

Due to
$\Pp^{b_1=0}_{\ref{F7}b}+\frac{1}{2}\Pp^{b_2=0}_{\ref{F7}c}=\Pp_{\ref{F7}e}$
we get
$2\Gamma_{\ref{F7}b}+\Gamma_{\ref{F7}c}+2\Gamma_{\ref{F7}e}=\Delta\Gamma^{(4)}_{2l}+(finite\:
terms)$.

%%%%%%%%%%%%%%%%%%%

}
\end{document}